\documentclass[12pt,twoside]{article}
\usepackage[dvips]{graphics,color}
\usepackage{array}
\newcommand{\ow}{\overline{W}}
\newcommand{\oz}{\overline{Z}}
\newcommand{\own}{\overline{w}}

\begin{document}
\rightline{KUCP0189}

\vspace{10mm}
\centerline{\Large\bfseries Small-World Effects in Wealth Distribution}

\vspace{20mm}
\centerline{\large Wataru Souma
\footnote{e-mail: souma@isd.atr.co.jp}}

\centerline{Information Sciences Division, ATR International,
Kyoto 619-0288, Japan}

\centerline{\large Yoshi Fujiwara
\footnote{e-mail: yfujiwar@crl.go.jp}}

\centerline{Keihanna Research Center, Communications Research
Laboratory, Kyoto 619-0289, Japan}

\centerline{\large Hideaki Aoyama
\footnote{e-mail: aoyama@phys.h.kyoto-u.ac.jp}}

\centerline{Faculty of Integrated Human Studies, Kyoto University,
Kyoto 606-8501, Japan}

\vspace{40mm}
\begin{abstract}
We construct a model of wealth distribution, based on
an interactive multiplicative stochastic process
on static complex networks.
Through numerical simulations we show 
that a decrease in the number of links discourages 
equality in wealth distribution, while the rewiring of links in small-world
networks encourages it.
Inequal distributions obey log-normal distributions, which
are produced by wealth clustering.
The rewiring of links breaks the wealth clustering and
makes wealth obey the mean field type (power law) distributions.
A mechanism that explains the appearance of log-normal distributions
with a power law tail is proposed.
\end{abstract}

\newpage
With the advance of the theory of complex networks \cite{WS},
it has become clear that many social networks
can be understood as small-world networks \cite{NSAB}.
On the other hand, recent high-precision studies
of economic phenomena, helped by the availability of 
high-quality data in digital format \cite{souma}, is starting 
to reveal the basic characterics of wealth distribution,
which form a basis and at the same time a final product of 
economic activities.
In view of these developments, we propose a stochastic model of
wealth distribution built on a complex network
and present the results of our numerical simulation in this paper.

Several models for wealth distribution 
have been proposed
with some reality and success \cite{IKR,solomon1,BM,BMetc}.
They, however, lack one feature we consider important:
In some models, each agent (economic body) goes through a stochastic
process of increasing and decreasing wealth, quite
independently from other agents.
In some others, each agent interacts with either randomly-selected
agents or neighboring agents on a fixed lattice.
Apparently, such a selection of economic partnership/competition
is far from reality, as is evident from complex network studies.
This is one reason we propose our model, which features both
the stochastic nature of economic activities and underlying
network structures that are expected to be close to
those of the relevant social activities.
We also note that studies of economic phenomena that build 
upon the underlying complex networks are novel.

In considering the mechanisms behind wealth distribution, 
it is important to include at least two basic natures of the wealth.
One is a random change of the value of wealth and
the other is the liquidity of it.
As a minimal model that contains these features 
we chose an interactive multiplicative stochastic process \cite{BM},
which is defined by the following evolution equation:
\begin{equation}
\label{eq:master}
W_i(t+1)=a_i(t)W_i(t)+\sigma^2W_i(t)+\sum_{j(\ne i)}J_{ij}(t)W_j(t)
-\sum_{j(\ne i)}J_{ji}(t)W_i(t),
\end{equation}
where $W_i(t)$ is the wealth of the $i$-th agent at the time $t$,
$a_i(t)$ is a Gaussian random variable with mean $m$ 
and variance $2\sigma^2$,
which describes the spontaneous increase or decrease of the wealth.
The rest of the terms in the r.h.s. of Eq.~(\ref{eq:master}) 
describe transactions on business networks.
We note that in Ref.~\cite{BM} 
the process corresponding to Eq.~(\ref{eq:master})
is described by a stochastic differential equation in the Stratonovich
sense, which generates the second term proportional 
to $\sigma^2$ when discretized.
In addition, although naive discretization brings in a
time interval $\Delta t$ in Eq.(\ref{eq:master}),
it has been absorbed by the rescaling of the parameters.

The transaction matrix $J_{ij}$ is chosen based on the underlying
network structure:
If the $i$-th agent is directly connected to the $j$-th agent by a link,
we take $J_{ij}=J/Z_i$,
where $Z_i$ is the number of agents connected to the $i$-th 
agent (``neighbors"). Otherwise, $J_{ij}=0$.
The parameter $J$ is a positive number, which 
represents the ratio of transactions and the wealth.
Thus, the third term in the r.h.s. of Eq.~(\ref{eq:master}) describes
the incoming wealth, and the fourth term describes the outgoing wealth.
Under this selection of transaction matrix, Eq.~(\ref{eq:master}) 
can be written as follows:
\begin{equation}
\label{eq:disc}
W_i(t+1)=\left\{a_i(t)+\sigma^2-J\right\}W_i(t)+J\ow_i(t),
\end{equation}
where $\ow_i(t)$ is the mean wealth of the neighbors of the $i$-th agent.

We consider one-dimensional and static network of $N$-agents 
with periodic boundary conditions \cite{WS,NSAB}, 
which are determined by two parameters, i.e., a rewiring
probability $p$ and a mean number of links per agent $\oz$.
(We will also use a notation $z\equiv \oz/N$.)
The rewiring probability $p$ is defined by
$p\equiv(\textrm{the number of rewired links})/
(\textrm{the total number of links})\in[0,1]$.
At $p=0$ we have a regular network with $Z_i=\oz$, in which the $i$-th agent is
connected only with $\{i-\oz/2, \cdots,i-1,i+1,i+\oz/2\}$-th agents.
For $p\ne0$, links are rewired to conserve the total number of links.
At $p=1$ we obtain completely random networks.

In the mean field (MF) case, i.e., $z\simeq1$,
the stationary probability density function $p(w)$
is known to be the following in the large $N$ limit \cite{BM},
\begin{equation}
\label{eq:mfsolution}
p(w)=\frac{(\mu-1)^\mu}{\Gamma[\mu]}\frac{\exp\{-(\mu-1)/w\}}{w^{1+\mu}},
\end{equation}
where $w_i(t)$ is the wealth normalized by the mean wealth
$\ow(t)$; $w_i(t)=W_i(t)/\ow(t)$.
(We will simply call $w_i(t)$ the normalized wealth hereafter.)
The exponent is given by $\mu=1+J/\sigma^2$.
In the large $w$ range Eq.~(\ref{eq:mfsolution}) becomes a power law
distribution, $p(w)\propto w^{-(1+\mu)}$, and therefore the exponent $\mu$ is
the Pareto index in economics.

\begin{figure}[h]
\begin{center}
\resizebox{.9\textwidth}{!}{\includegraphics{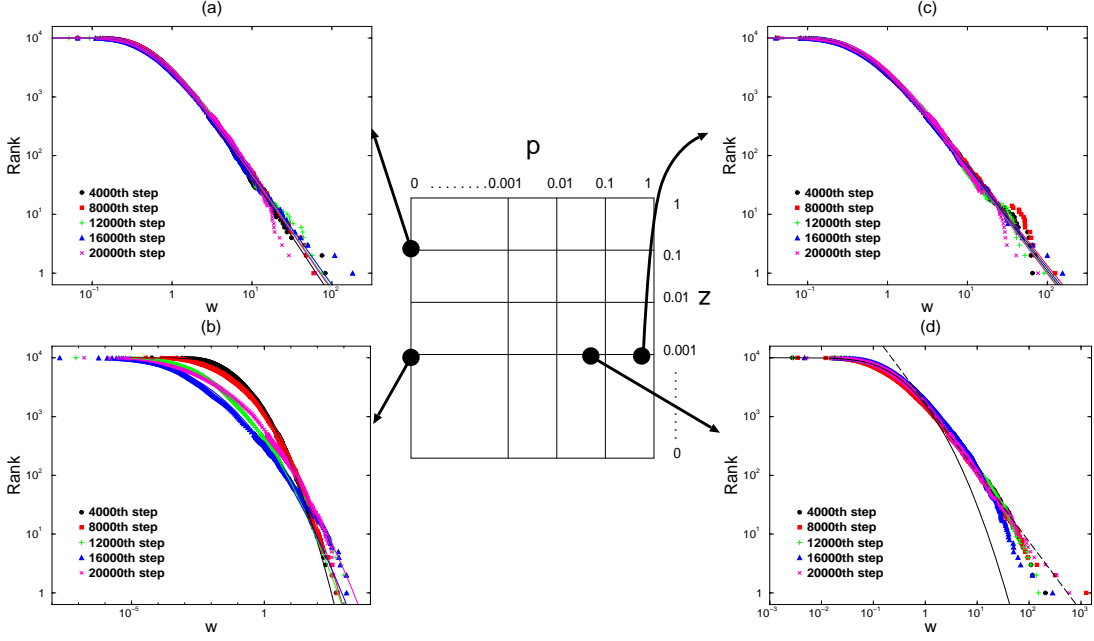}}
\end{center}
\caption{\label{fig:1}
Some of the wealth distributions obtained by numerical simulations.
The rank distribution in regular networks is plotted in (a) for $z=0.1$
and in (b) for $z=0.001$.
The rank distribution in small-world networks 
is plotted in (c) for $z=0.001$ and $p=0.01$ 
and in (d) for $z=0.001$ and $p=0.5$.}
\end{figure}

We have carried out numerical simulations for $N=10^4$,
$m=1$, $J=0.01$ and $\sigma^2=0.01$.
Under this parameter set, the Pareto index in the MF case is
$\mu_{\textrm{mf}}=2$.
Some of the distributions we obtained are shown in Figs.~\ref{fig:1}~(a)-(d).
In these figures the horizontal axis is the logarithm of
$w$ and the vertical axis is that of the rank.
We have numerically calculated up to $2\times 10^4$ time steps,
and drawn distributions at each $4\times 10^3$ step.
The rank distributions in the case of regular networks
with $z=0.1$ correspond to Fig.~\ref{fig:1}~(a).
The solid lines show the best fits with the MF
result Eq.~(\ref{eq:mfsolution}).
We see that all of the resulting distributions fit well with
the MF result with $\mu\simeq\mu_{\textrm{mf}}$.
Figure~\ref{fig:1}~(b) is a case of regular networks with $z=0.001$.
We find that the results do not agree with the MF result,
but instead fit well with log-normal distributions,
which are shown by the solid lines.
Note that the best fit log-normal distributions have
time-dependent mean value $m_{\textrm{ln}}(t)$ and
time-dependent variance $\sigma^2_{\textrm{ln}}(t)$.
On the other hand, if we rewire links with high probability
the MF type distributions reappear.
Figure~\ref{fig:1}~(c) is a case of regular networks
with $z=0.001$ and $p=0.5$.
The solid lines are the fitting by
MF type solutions with $\mu=1.7\sim 1.8$, which is slightly
smaller than $\mu_{\textrm{mf}}$.
Hence we observe that the change in wealth distribution occurs 
at the intermediate value of $p$.
Figure~\ref{fig:1}~(d) is a case of small-world networks
with $z=0.001$ and $p=0.05$.
We have found that the distributions can not be fitted well by either the
log-normal functions or the MF type functions.
Interestingly, however, we find that
a combination of log-normal and power-law distributions fits
well with the results.
One such example at the $2\times 10^4$-th time step
is shown with a log-normal function
for middle and low wealth ranges (solid line)
and a power-law function for the high wealth range (dashed line).
Log-normal distributions with a power law tail like this are
frequently observed
in real world economic phenomena \cite{souma}.

\begin{figure}[h]
\begin{center}
\resizebox{0.45\textwidth}{!}{\includegraphics{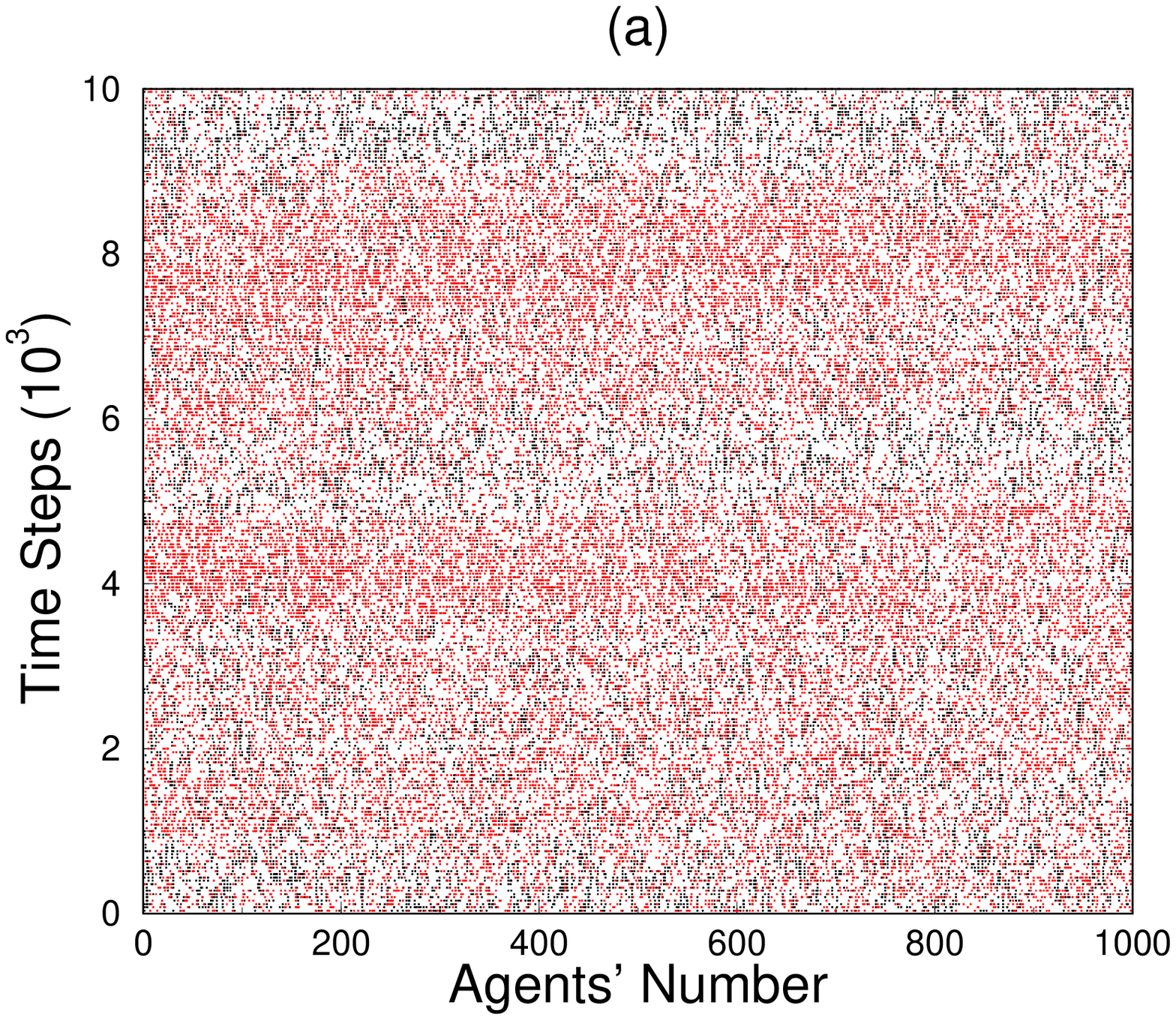}}
\resizebox{0.45\textwidth}{!}{\includegraphics{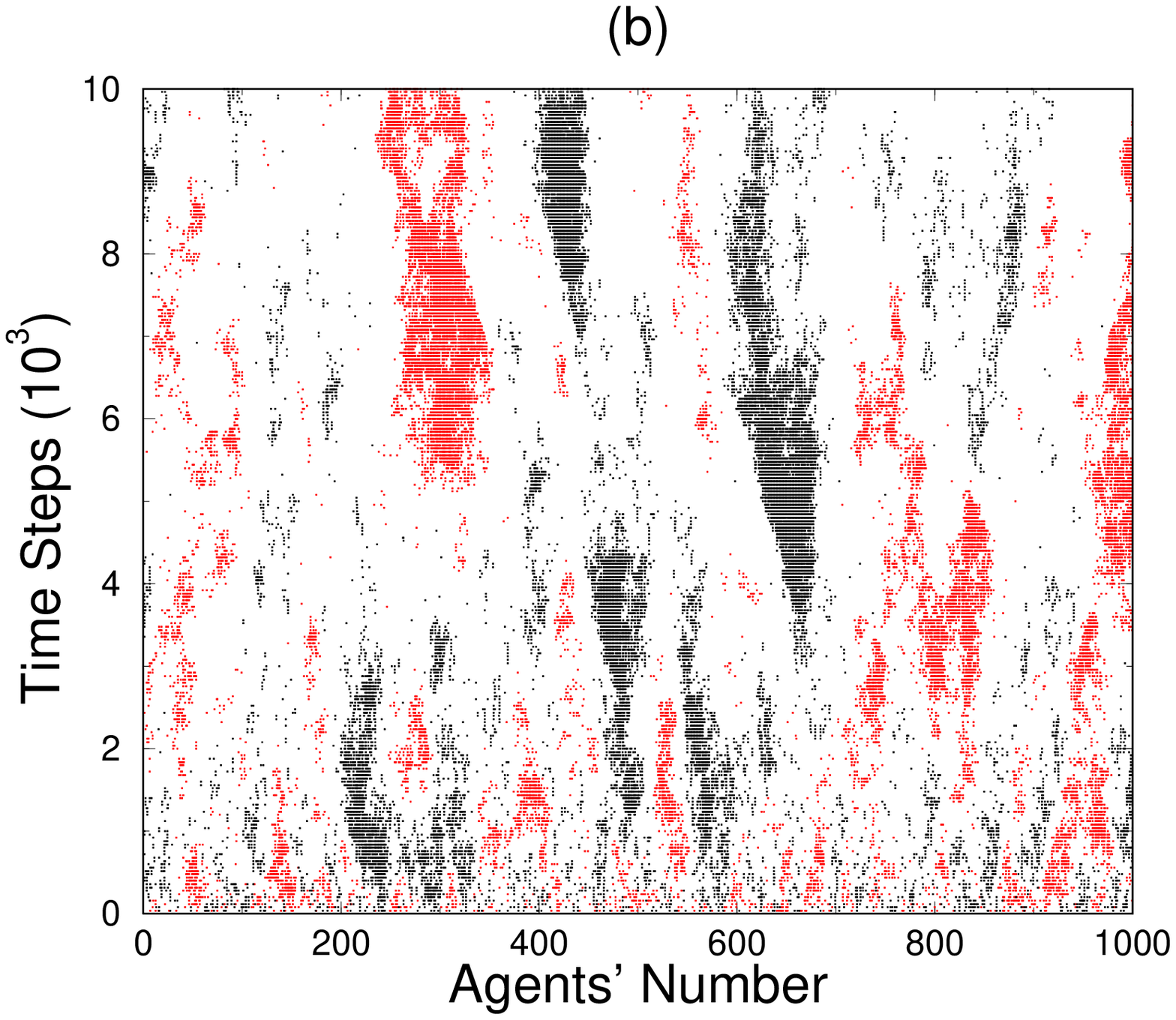}}\\
\resizebox{0.45\textwidth}{!}{\includegraphics{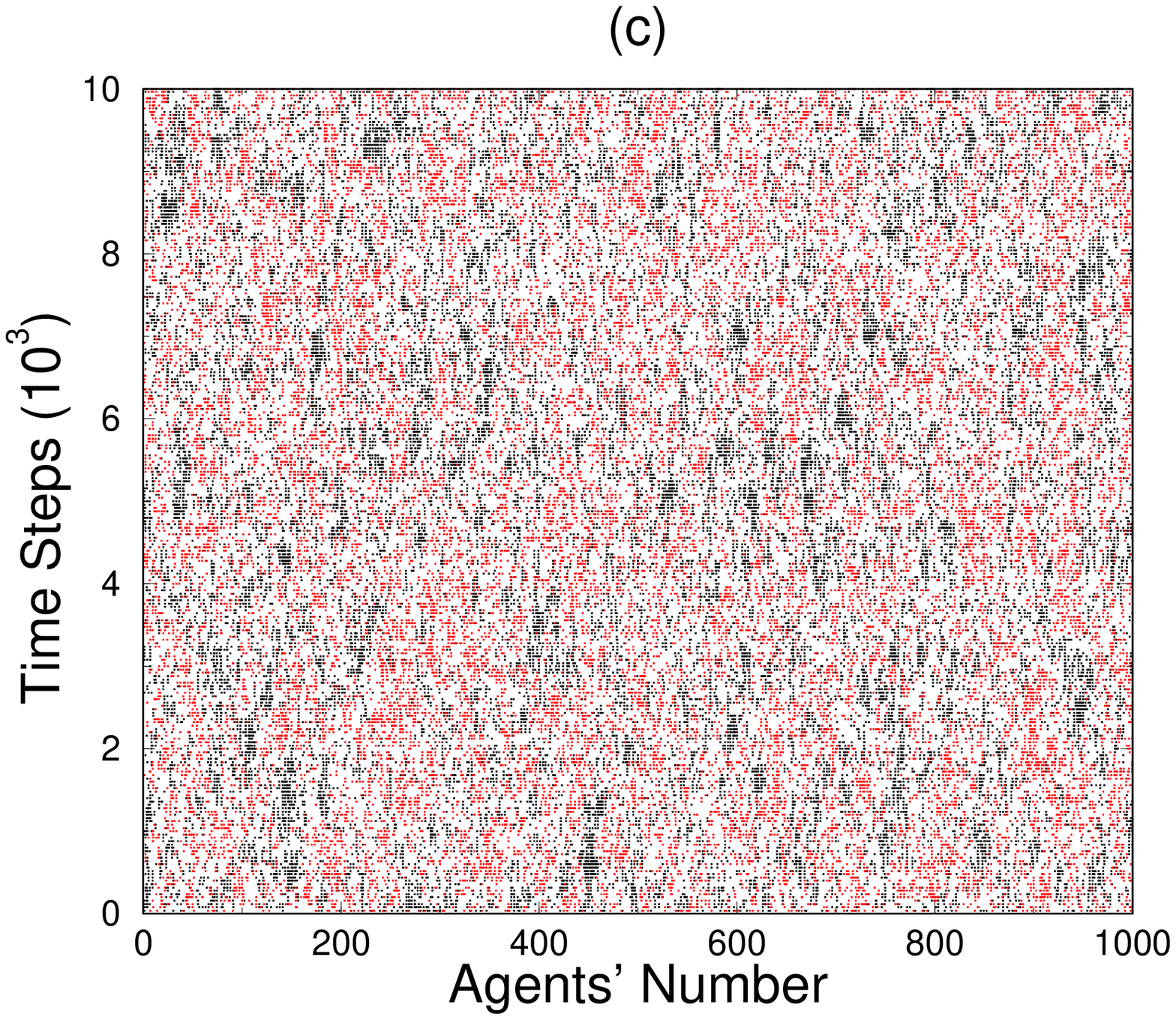}}
\resizebox{0.45\textwidth}{!}{\includegraphics{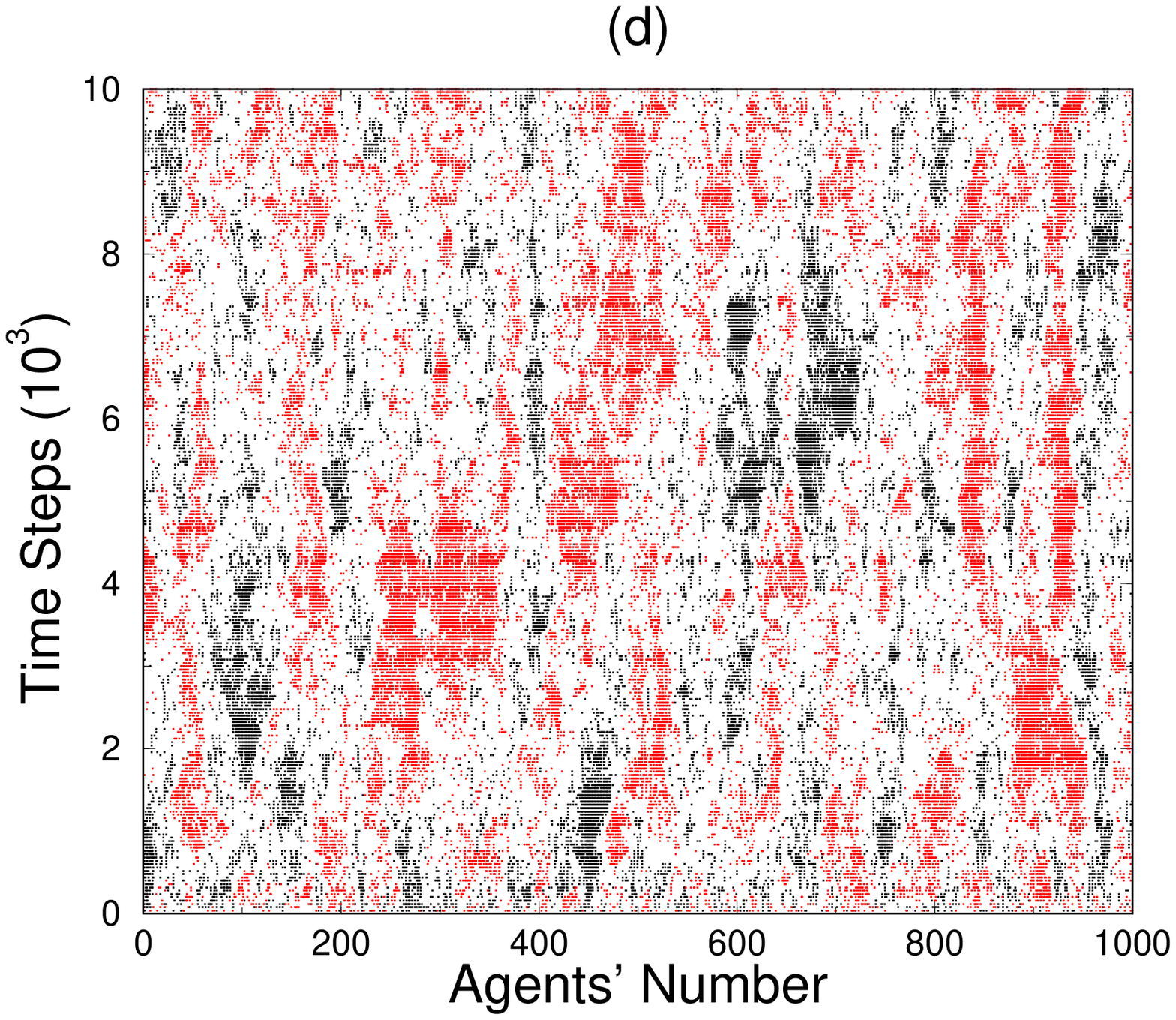}}
\end{center}
\caption{\label{fig:2}
Wealth developments for regular networks with $z=0.1$ are shown in (a)
and with $z=0.001$ in (b).
The case of small-world networks with $z=0.001$
and $p=0.05$ is shown in (c), and $p=0.5$  in (d).}
\end{figure}

It is natural to question why distribution patterns change.
To answer this question, we directly consider the development
of the wealth.
Examples of numerical simulations
are shown in Figs.~\ref{fig:2}~(a)-(d).
In these figures the horizontal axis shows the agents' number
and the vertical axis shows time steps.
Though we have simulated for $N=10^4$ agents, wealth developments
for $10^3$ agents are shown.
We have calculated up to $10^4$ time steps.
If any agent ranks in the top $10\%$, we mark it with a black dot, and
if any agent ranks in the bottom $10\%$ we mark it with a red dot.

Figure~\ref{fig:2}~(a) is a case of regular networks with $z=0.1$.
We can see that the black and red dots are uniformly distributed
without notable
clustering.
Since each agent has $10^3$ neighbors,
each agent transacts with agents from various ranks, covering a wide range.
Figure~\ref{fig:2}~(b) is a case of regular networks with $z=0.001$.
Contrary to the case of $z=0.1$, the black and red dots cluster heavily.
This ``wealth clustering'' phenomenon is supported by the fact that
each agent has only $10$ neighbors and rich agents transact mainly 
with rich agents and poor agents transact mainly with poor agents.
Hence $W_i(t)$ and $\ow_i(t)$ have strong correlations.
Actually, for the case of regular networks
with $z=0.001$, we can obtain
$\langle\own_i(t)w_i(t)\rangle=0.92\sim0.95$.
Here $\own_i(t)\equiv\ow_i(t)/\ow(t)$.
On the other hand
we can obtain $\langle\own_i(t)w_i(t)\rangle=0.02\sim0.09$
for the case of regular networks with $z=0.1$.
Figure~\ref{fig:2}~(c) is a case of small-world networks
with $z=0.001$ and $p=0.5$.
This case is almost similar to Fig.~\ref{fig:2}~(a).
Figure~\ref{fig:2}~(d) is 
a case of small-world networks
with $z=0.001$ and $p=0.05$.
Although wealth clustering is observed, the size of each cluster
is smaller than the case of regular networks with $z=0.001$.
From these results we find that
a decrease in the number of links causes wealth clustering and
the rewiring of links disperses it.

If $W_i(t)$ and $\ow_i(t)$ have strong correlations,
Eq.~(\ref{eq:disc}) becomes a
pure multiplicative stochastic process, and the wealth obeys
log-normal distributions with mean $vt$ and variance $Dt$,
where
$v=\langle\log\{a(t)+\sigma^2\}\rangle$ and
$D=\langle[\log\{a(t)+\sigma^2\}]^2\rangle-
\langle\log\{a(t)+\sigma^2\}\rangle^2$.
On the other hand if $W_i(t)$ and $\ow_i(t)$ have no correlations,
Eq.~(\ref{eq:disc}) is regarded as a multiplicative stochastic
process with additive noise \cite{kesten}.
It is known that this process induces
power law distributions in large $W_i(t)$ ranges,
if two conditions are satisfied:
One condition is that the stochastic variable $a_i(t)$ and
the additive noise $J\ow_i(t)$ must behave independently.
The other condition is
$\langle\log\{a_i(t)+\sigma^2-J\}\rangle<0$.
Although the first condition does not matter in our model,
the parameter ranges of
$J$ and $\sigma^2$ are restricted by the second condition.
If $W_i(t)$ and $\ow_i(t)$ have no correlations
in the high wealth ranges and have strong correlations in the 
middle and low wealth ranges, the wealth will obey log-normal 
distributions with a
power law tail such as those observed in many real world economic phenomena
\cite{souma}.

\begin{figure}[t]
\begin{center}
\resizebox{0.45\textwidth}{!}{\includegraphics{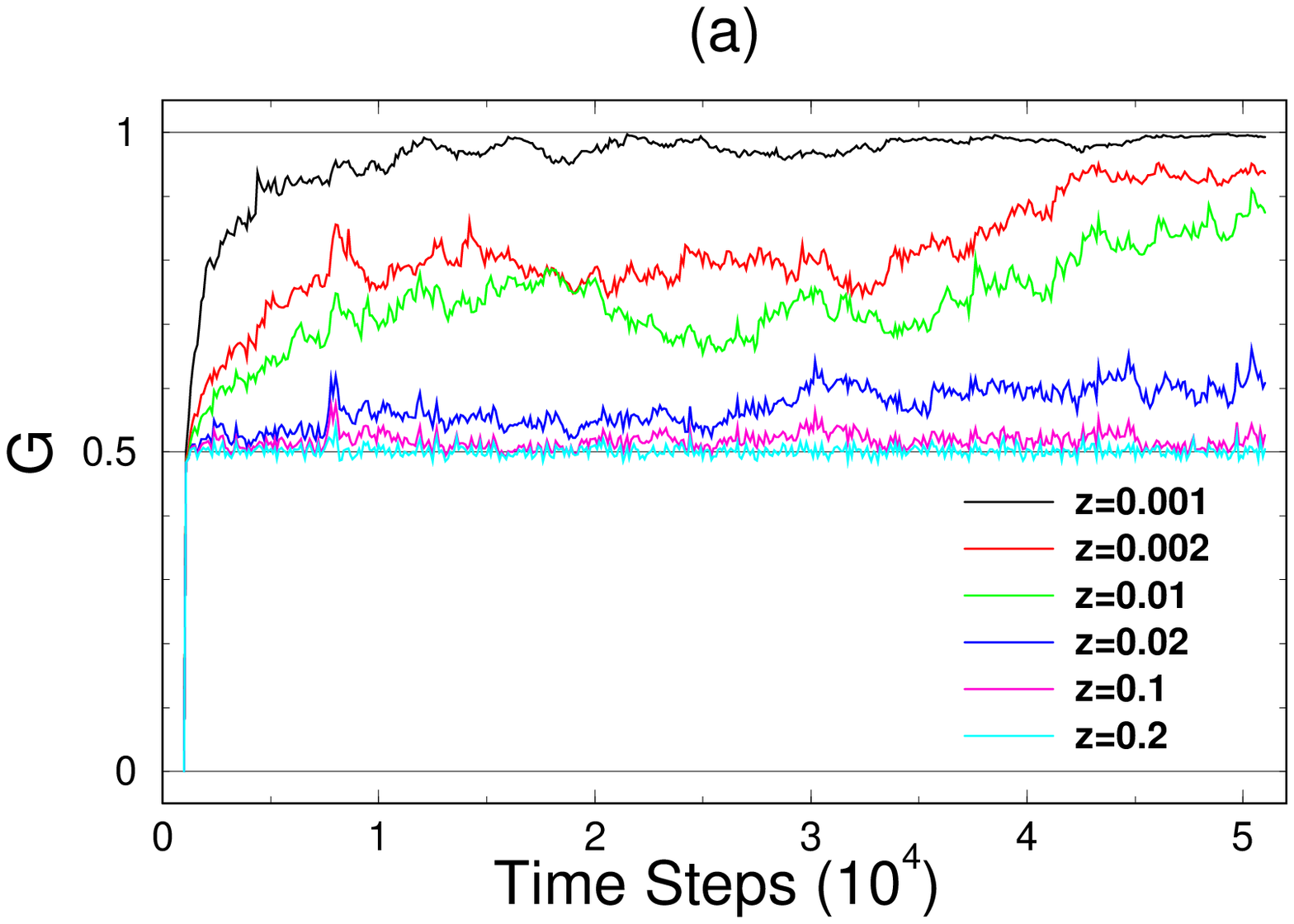}}
\resizebox{0.45\textwidth}{!}{\includegraphics{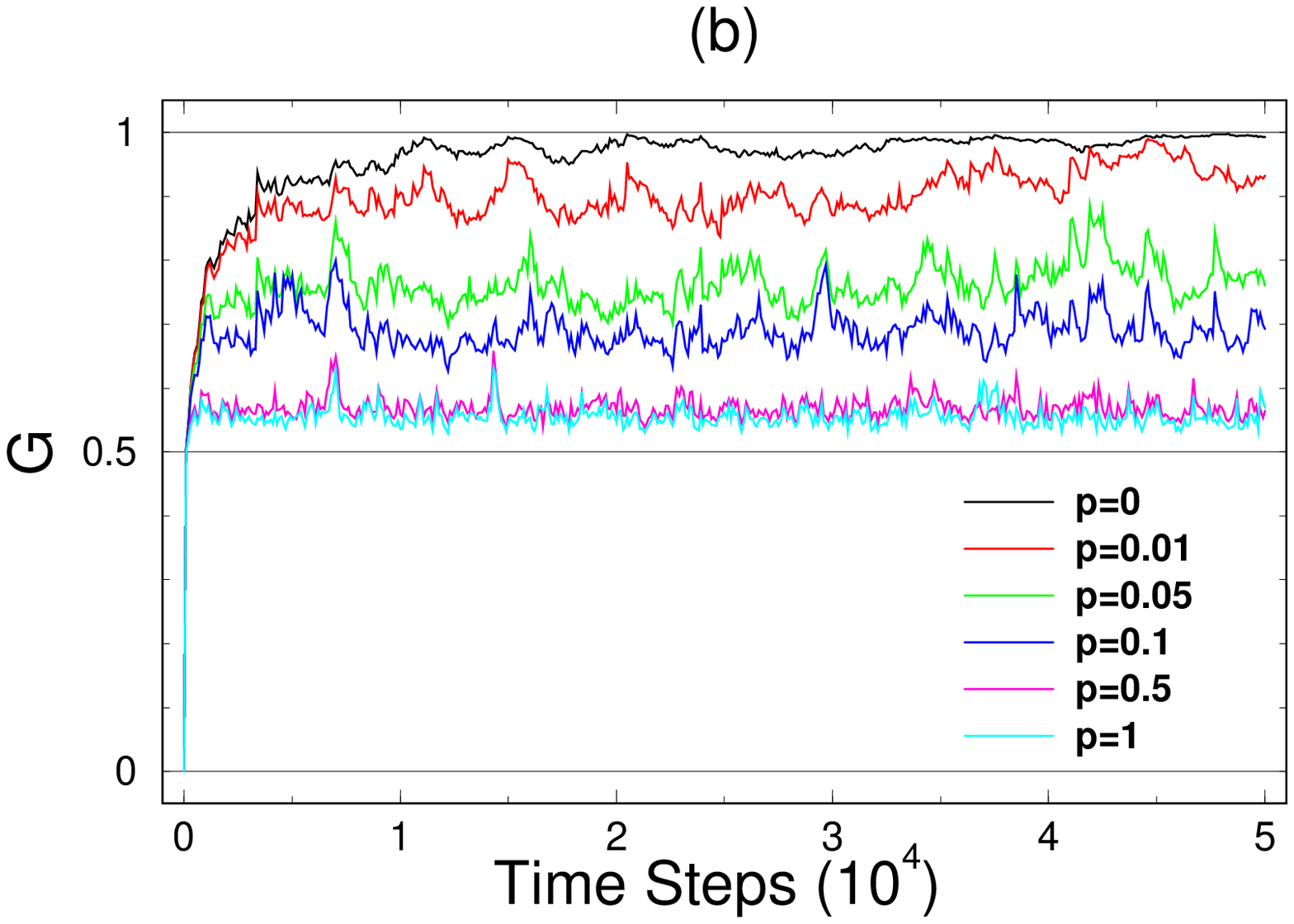}}
\end{center}
\caption{\label{fig:3}
Change of Gini coefficient in the case of regular networks (a),
and that of small-world networks (b).
}
\end{figure}
We will now study the inequality of wealth distribution
by using the Gini coefficient $G$ to quantify it.
Take
every possible pair of income recipients (wealth possessors) and
calculate the average of the absolute differences of the two incomes
(wealth) for each pair. $G$ is then half the ratio of the average to
the mean value of all the incomes (wealth). Obviously $G=0$ when
everyone has the same, and $G=1$ when the entire amount is
concentrated in a single person.
With our parameter set, the Gini coefficient of the MF
case is $G_{\textrm{mf}}=0.5$.

The temporal changes of $G$ are shown in Figs.~\ref{fig:3} (a) and (b).
In these figures, the horizontal axis shows time steps
and the vertical axis shows the value of $G$.
We have numerically simulated up to $5\times 10^4$ time steps.
The temporal changes of $G$ in the case of 
regular networks are shown in Fig.~\ref{fig:3}~(a).
In this figure we observe
that $G$ gradually increases with time steps
in the case of $z=0.01$ and $0.02$, which is
in accordance with the log-normal distributions with $m_{\textrm{ln}}(t)$
and $\sigma^2_{\textrm{ln}}(t)$.
We also observe that almost completely inequal distributions
appear in the case of $z=0.001$.
We find that the decrease of $z$ makes distributions inequal.
The temporal changes of $G$ in the case of
small-world networks with $z=0.001$ are shown in Fig.~\ref{fig:3}~(b).
From this figure we find that the rewiring of links in
small-world networks makes distributions equal.

\begin{figure}[t]
\begin{center}
\resizebox{0.5\textwidth}{!}{\includegraphics{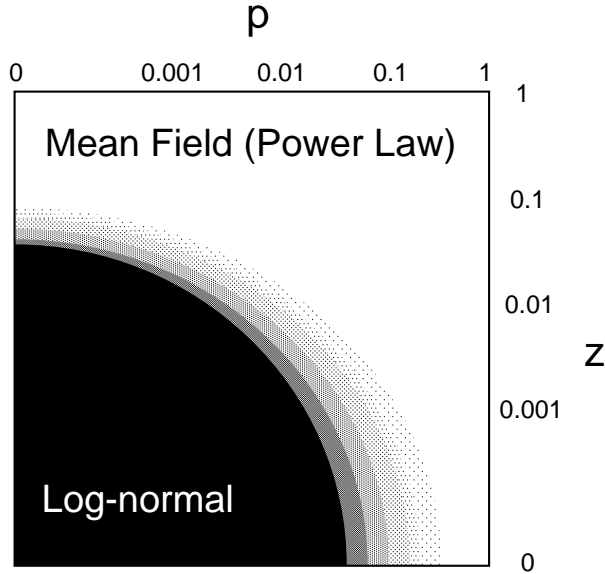}}
\end{center}
\caption{\label{fig:4}
Patterns of wealth distribution in the case of $N=10^4$.
}
\end{figure}

The results of this study are summarized in two parts.
First, we have carried out a simulation
for a wide range of values of $p$ and $z$,
which is summarized by the phase diagram shown in Fig.~\ref{fig:4}.
This diagram has two extreme distributions,
the log-normal distributions with
$m_{\textrm{ln}}(t)$ and $\sigma^2_{\textrm{ln}}(t)$
and the MF type distributions represented by Eq.~(\ref{eq:mfsolution}).
These two ranges are interpolated by 
a decrease in the number of links or the rewiring of the links.
In the intermediate region, the wealth distribution may be fitted well
with log-normal distributions with a power law tail,
which are observed in real world economic phenomena \cite{souma}.
Lastly, from the study of the Gini coefficient 
we have found that a decrease in the number of links
makes wealth distribution inequal and the rewiring of links in
small-world networks makes it equal.
Hence society will be protected against exceeding inequalities
in wealth by small-world effects in economic networks.

We note that it would be very interesting to carry out an analysis of 
models similar to the ones studied here, but built on other kinds of
networks. For example, the network can be dynamic, allowing 
rewiring to be either independent from or dependent on wealth.
Other novel networks, like scale-independent networks,
would also be worth studying.
We believe that the work presented here forms a basis for the
study of general stochastic processes on complex networks.

The authors would like to thank Dr.~K.~Shimohara (ATR-ISD)
for his continuous encouragement and warm support.

\end{document}